\documentclass{article}

\usepackage[margin=3cm]{geometry}
\usepackage{graphicx}
\usepackage{booktabs}
\usepackage{multirow}
\usepackage{subfig}
\usepackage[round]{natbib}
\usepackage{amsmath}
\usepackage{enumerate}
\usepackage{comment}
\usepackage[usenames]{color}
\usepackage{threeparttable}
\usepackage{url}
\usepackage{lineno}

\newcommand{\degrees}{\ensuremath{^{\circ}}}
\newcommand{\mr}{\multirow}
\newcommand{\mc}{\multicolumn}
\newcommand{\tej}{Tropical Easterly Jet}
\newcommand{\cam}{CAM-3.1}
\newcommand{\nglo}{noGlOrog}
\newcommand{\ap}{aqua-planet}
\newcommand{\fp}{full-physics}
\newcommand{\ip}{ideal-physics}
\newcommand{\um}{$U_{max}$}
\newcommand{\un}{$U_{norm}$}
\newcommand{\ms}{ m s$^{-1}$}
\newcommand{\kd}{ K day$^{-1}$}
\newcommand{\td}{$\tau_{d}$}
\newcommand{\dc}{\degrees{}C}
\newcommand{\de}{\degrees{}E}
\newcommand{\dw}{\degrees{}W}
\newcommand{\dn}{\degrees{}N}
\newcommand{\ds}{\degrees{}S}
\newcommand{\ut}{upper tropospheric friction}

\begin{document}

\title{The impact of \ut{} and Gill-type heating on the location and strength of the Tropical Easterly Jet: Idealized physics in a dry Atmospheric General Circulation Model}

\author{S. Rao$^{1,2}$ \\
           $^1$Department of Mechanical Engineering \\
               Indian Institute of Science \\
               Bangalore 560012, India \\
           $^2$Engineering Mechanics Unit \\
               Jawaharlal Centre for Advanced Scientific Research \\
               Bangalore 560064, India \\
               Email: {samrat.rao@gmail.com, samrat.rao@jncasr.ac.in} \\
               Mobile: +919916675221}

\maketitle

\begin{abstract}
An atmospheric general circulation model (AGCM) with idealized and complete physics has been used to evaluate the \tej{} (TEJ) jet. In idealized physics, the role of \ut{} has been found to be important in getting realistic upper tropospheric zonal wind patterns in response to heating. In idealized physics, the location and strength of the TEJ as a response to Gill heating has been studied. Though the Gill model is considered to be widely successful in capturing the lower tropospheric response, it is found to be inadequate in explaining the location and strength of the upper level TEJ. Heating from the Gill model and realistic \ut{} does not lead to the formation of a TEJ. \\ \\
\textbf{Keywords:} \tej{}, Gill model, upper tropospheric friction
\end{abstract}

Short title: Upper tropospheric friction and location of the \tej{} in an idealized AGCM

\section{Introduction}\label{intro}

The \tej{} (TEJ) is an important upper tropospheric phenomena that occurs during the Indian Summer Monsoon Season (\cite{koteswaram-58}, \cite{flohn-68}). It has been shown that its location, structure and strength depend on the location and magnitude of heat sources (\cite{srao-13,srao-15}). But the methodology of using an atmospheric general circulation model (AGCM) has its own complications because of the detailed physics and parametrizations. \cite{gill-80}, henceforth G80, showed in his classic paper that heat sources in the tropics directly influence lower tropospheric wind and that the patterns bore qualitative resemblances to observations. It is from his paper that we know that a jet-like structure is possible even in a simple setting.

One of the important drawbacks of this model is that it is essentially a linear single-layer shallow water model, but actually represents a two layer baroclinic system. The upper and lower level winds are viewed as a single response. The zonal wind, which was a response to diabatic heating, was a cosine function of height which caused it to reverse signs in the lower and upper troposphere with maxima at top and bottom. Both upper level heating and boundary layer convergence are thus tightly coupled. It thus ignores the role of latent heating in driving the surface winds, which is true only if heating is at much higher levels (\cite{neelin-89}).

The requirement of a strong equivalent linear mechanical damping to get realistic wind response is also a consistent critique of this model. The frictional parametrization is identical in the lower and upper troposphere. The source of this damping in the upper troposphere is not clear. \cite{lin-08} showed that the strong damping used in Matsuno-Gill-type models was spatially inhomogeneous, and depended on the heating rate.  as prescribed by the deep convective or shallow clouds. \cite{gregory-97} found that the \tej{} in the Met Office (UKMO) GCM was weakened due to damping caused by Convective Momentum Transport parametrization. \cite{showman-10} showed that in Matsuno-Gill models equatorial easterlies persist due to the presence of momentum transport by the imposed heat source. \cite{hoskins-95} used a primitive equation model that used Rayleigh damping which increased as one moved into the stratosphere and beyond. The use of a multi-layered atmosphere model for studying the TEJ is not new. \cite{mishra-83} showed the importance of beta effect for wave growth of the TEJ. The effects of non-linearity on the TEJ was also studied. \cite{mishra-83,mishra-87} further showed that non-linearity was important in making the jet have a longer zonal extent.

In this paper, an AGCM, the Community Atmosphere Model, version-3.1 (\cam{}) has been used to study the effect of \ut{} on the TEJ. Heating as specified from the \cite{gill-80} model has been applied to study the location and structure of the TEJ. The physics in \cam{} has been considerably simplified by totally removing the entire default physics parametrizations. Instead the model has been run with specifications as in \cite{held-94} (hereafter referred to as HS94), unless explicitly stated. Such simulations with idealized physics have been termed as `\ip{}'.

\section{Model description and experimental details} \label{simdetails}

The AGCM that has been used is \cam{}. The finite-volume (FV) dynamical core with 2\degrees{}$\times{}$2.5\degrees{} horizontal resolution and 26 vertical levels, has been used. The time step is the default value of 30 minutes. The physics is as in HS94, unless explicitly stated. Since \cam{} has hybrid-sigma co-ordinates, frictional damping from levels 1 to $\sim$0.7 is identical to HS94. Newtonian relaxation to a specified equilibrium temperature is also identical to HS94.

The effect of \ut{} has been explored by comparing simulations with and without \ut{}, against simulations with full AGCM physics, henceforth called \fp{} simulations. Simulations had been conducted with the entire physics of \cam{}. The actual heating (inclusive of both convective and radiative) and temperatures from these simulations have been prescribed in the \ip{} mode of \cam{}. Then the impact of \ut{} has been assessed by comparing the location and strength of the jet in \ip{} simulations against simulations with full AGCM physics.

The \fp{} simulations considered  are:
\begin{enumerate}
\item Aqua-planet simulation with a single heat source. Precipitation was induced by elevating temperatures with a peak value at 90\de{},20\dn{}. The peak value was 32\dc{} which was brought down to a value of 25\dc{}. The elevated SSTs had an oval shape with a 90\degrees{} major-axis as 16\degrees{} minor-axis. This simulation, named AP\_90e20n, has been run for twelve months and the data of the 12$^{th}$ month has been used.
\item A one year simulation of the default configuration of \cam{} with climatological SSTs, but without orography. Since the TEJ is the strongest in the month of July (\cite{srao-13}), the month of July has been considered. This simulation has been named \nglo{}.
\end{enumerate}

The names of the corresponding \ip{} simulations has `\_ip' appended to the original names. The inclusion of \ut{} is understood by further appending `\_nf' (no friction) or `\_pf' (friction present). For example, for the \ap{} simulation with \ut{}, the name for the corresponding \ip{} simulation is AP\_90e20n\_ip\_pf. The temperatures from the \fp{} simulation have been zonally averaged and then imposed in the \ip{} cases. Rayleigh friction is imposed in the \ut{}. Damping, which is a function of height, was initiated at a hybrid level of of 0.3. A 2$^{nd}$ degree polynomial was chosen to fit values at three data points: (i) damping timescale of 30 days at 0.25 hybrid level, (ii) damping timescale of 300 days at 0.085 hybrid level and (iii) damping timescale of 1000 days at 0.0 level. Table \ref{tab: sims} gives details about the simulations. In the 3$^{rd}$ column, \cam{} implies that frictional parametrization is the default as in \fp{} of \cam{}. The \ip{} simulations have been run for 12 months and the last six months  have been averaged to study the results. 

\section{Effect of \ut{}} \label{fric}

Before proceeding, a brief description of some terms are given. $Q_{max}$ denotes the maximum heating rate. \um{} denotes the value of peak zonal easterly and \un{} denotes the zonal velocities normalized by \um{}. In the figures \un{} contours are shown to better facilitate comparison between \fp{} and \ip{} simulations. Since the winds are normalized by \um{}, the positive contours in the figures are actually easterlies, and the negative contours are westerlies. Jet length and width are defined on basis of the location of \um{} and closure of the \un{} = 0.5 contour, that is, closure of the 0.5\um{} contour. Arcs in zonal and meridional directions passing through the location of \um{} are considered. Zonally and meridionally, the length in degrees when these arcs touch the 0.5\um{} contour are considered as jet length and widths respectively. If the 0.5\um{} contour does not close zonally, then the jet is defined to be non-existent.

\subsection{Aqua-planet simulation} \label{aqua}

\begin{figure}[htbp]
   \begin{center}
   \subfloat[AP\_90e20n (SST: 32\dc{} peak on 25\dc{} background)]{\label{ap-qmax-3225-xy}\includegraphics[trim = 5mm 20mm 0mm 60mm, clip, scale=0.35]{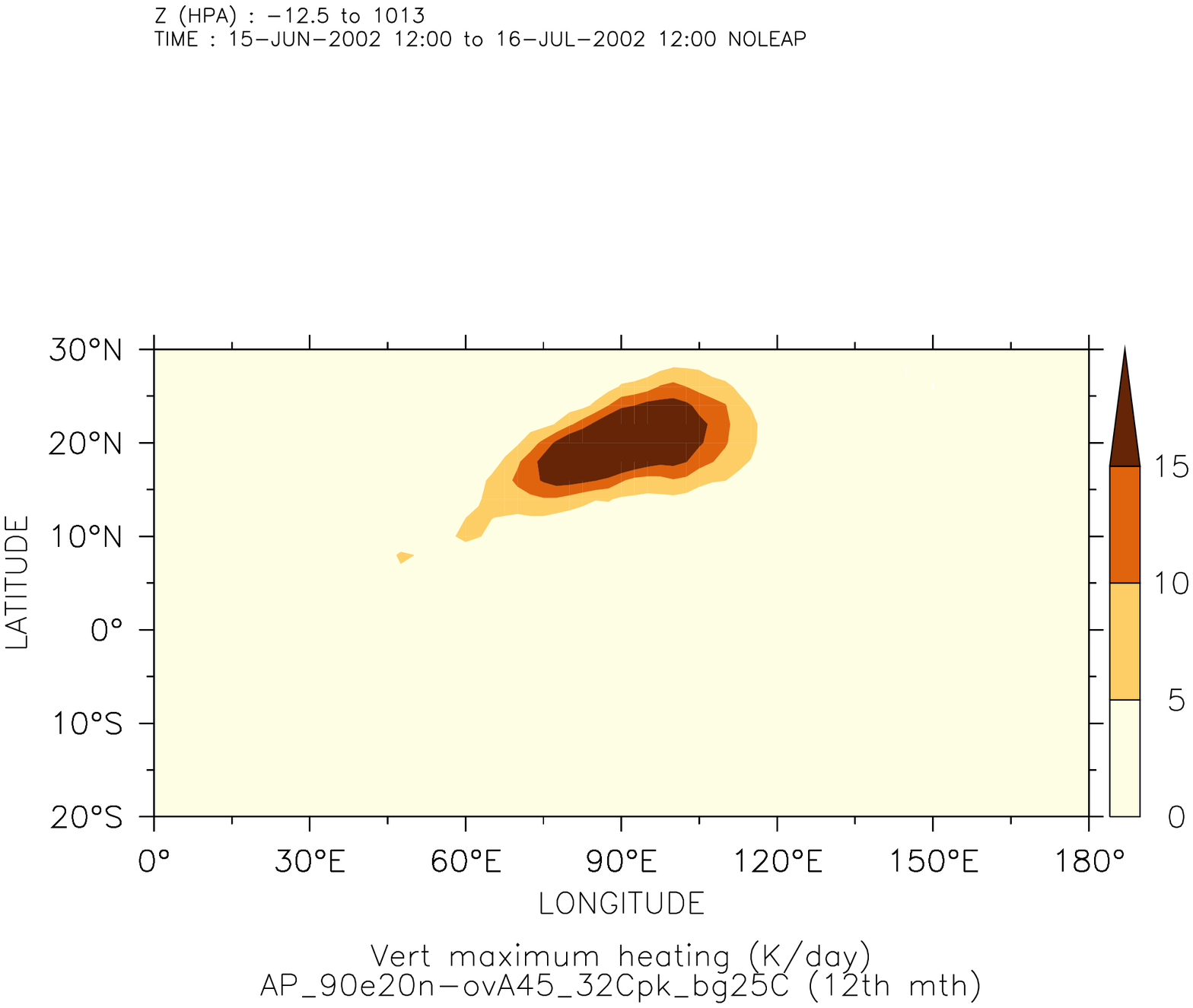}}
   \hspace{5mm}
   \subfloat[\nglo{} (no orography)]{\label{nglo-qmax-xy}\includegraphics[trim = 5mm 20mm 0mm 60mm, clip, scale=0.35]{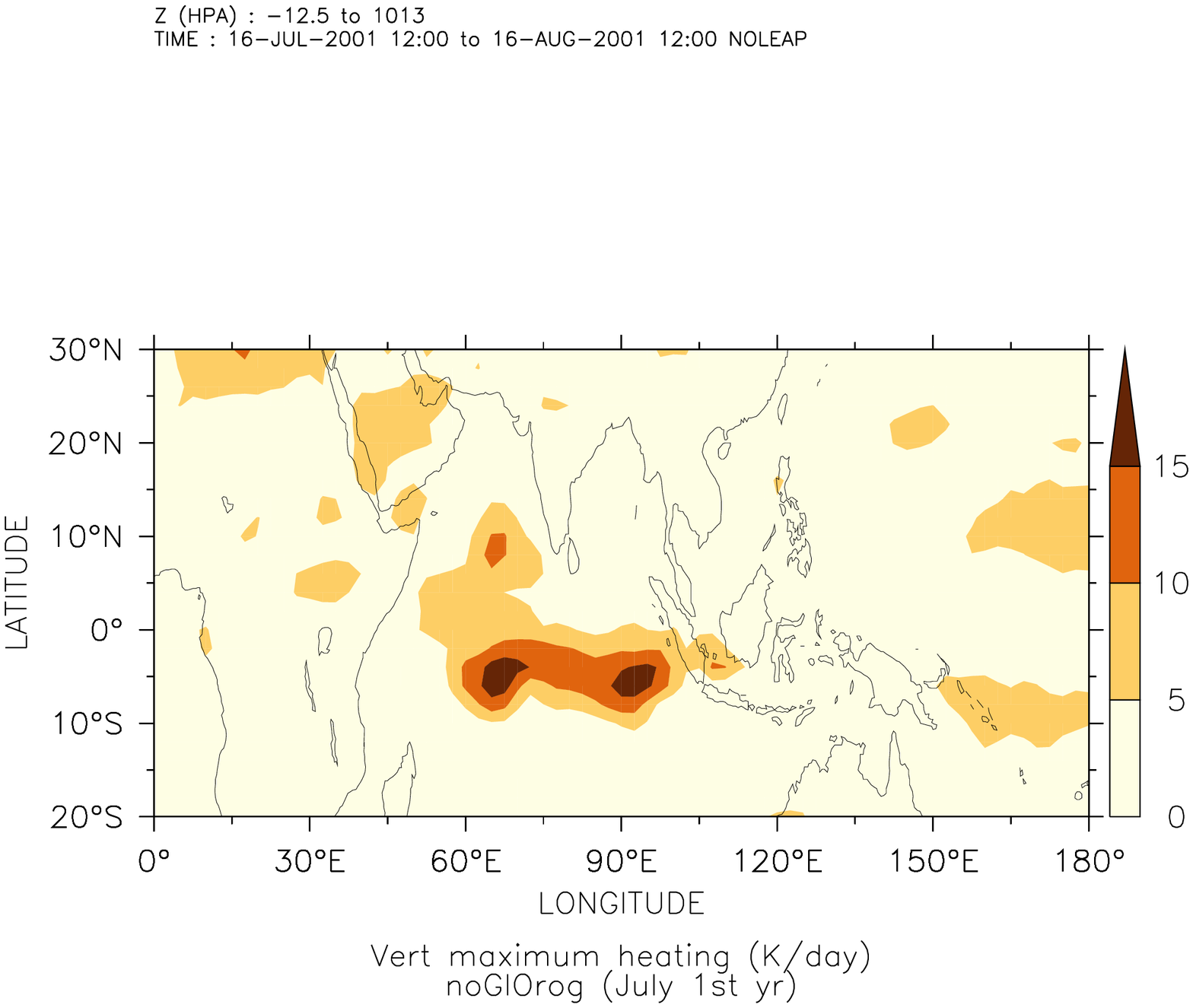}}
   \end{center}
   \caption{Maximum heating (\kd{}) rate of AP\_90e20n and \nglo{} simulation}
   \label{fig: nglo-ap-qmax-xy}
\end{figure}

The \ap{} configuration used has been described in section \ref{intro} above. The maximum heating rates is shown in Fig. \ref{ap-qmax-3225-xy}. The maximum heating has been shown because this enables easy understanding of peak heating, since at different locations the rates peak at different pressure levels . Table \ref{tab: sims} has details about the \fp{} and corresponding \ip{} simulations.

\begin{figure}[htbp]
   \begin{center}
   \subfloat[AP\_90e20n: \fp{}]{\label{ap-Unorm-3225-xy}\includegraphics[trim = 5mm 20mm 10mm 60mm, clip, scale=0.35]{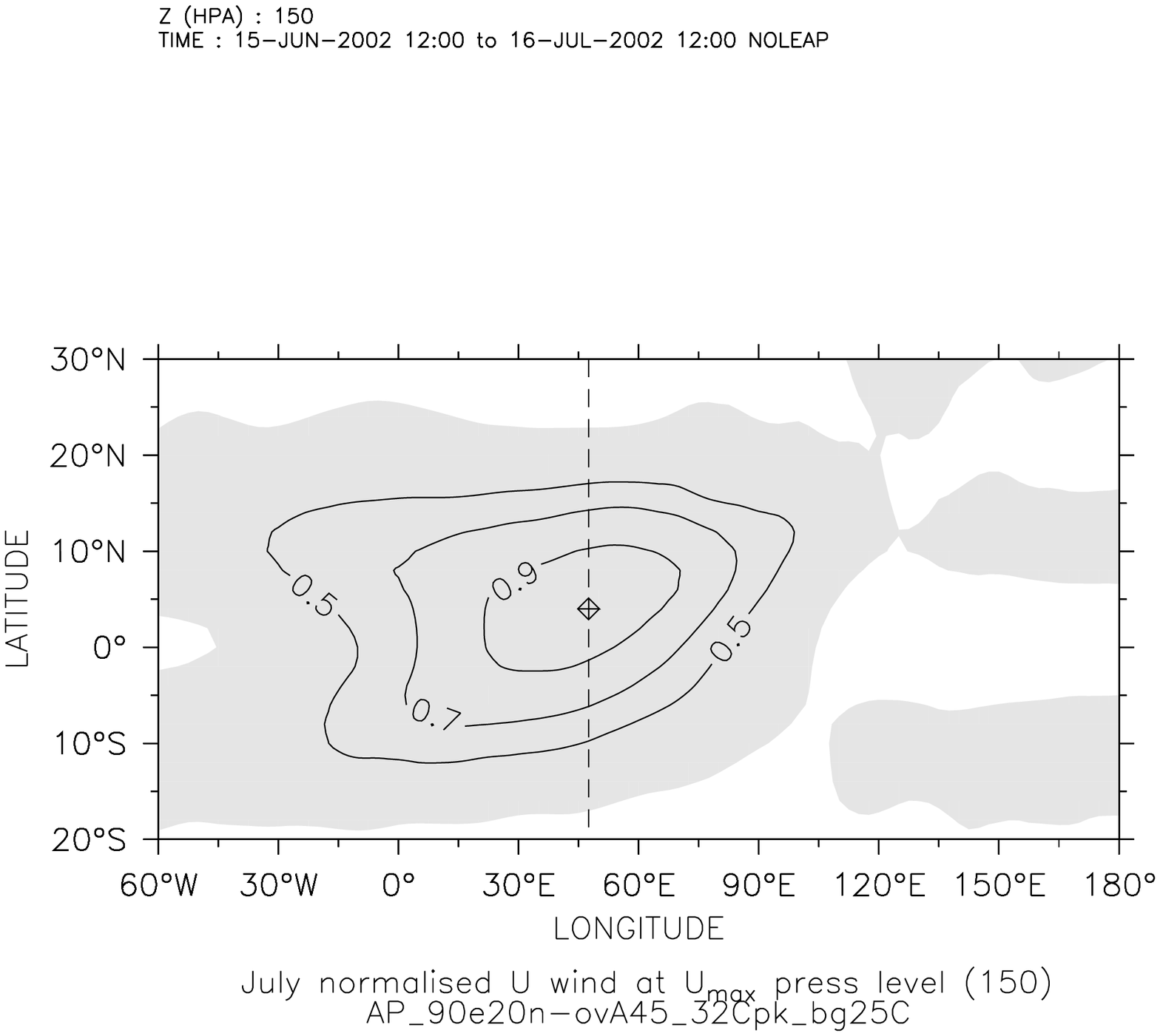}}
   \hspace{5mm}
   \subfloat[\nglo{}: \fp{}]{\label{nglo-Unorm-xy}\includegraphics[trim = 5mm 20mm 10mm 60mm, clip, scale=0.35]{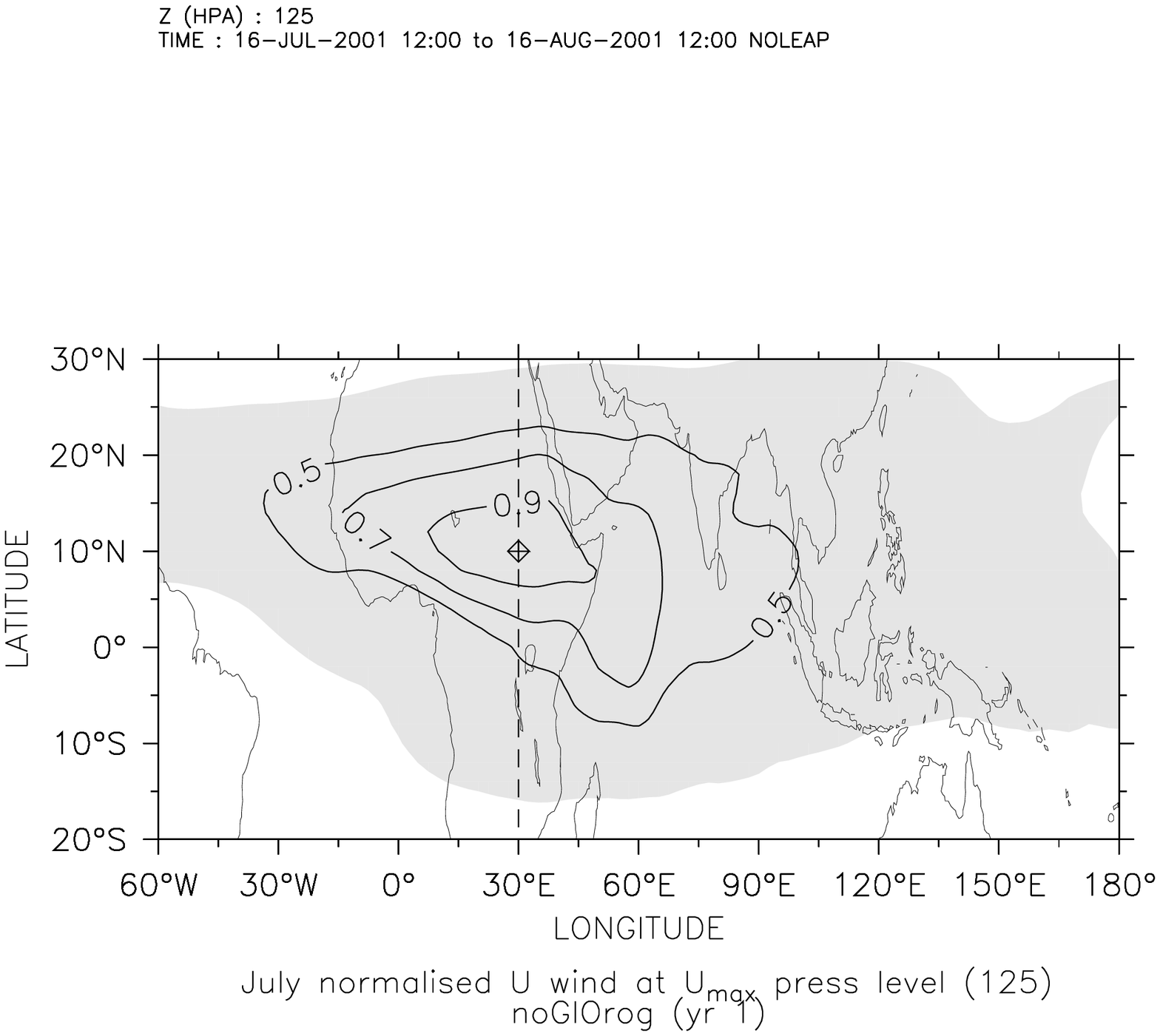}}
   \vskip 5mm
   \subfloat[AP\_90e20n\_ip\_nf: \ip{}]{\label{ap-cam3TeqXavg-Unorm-3225-xy}\includegraphics[trim = 5mm 20mm 10mm 60mm, clip, scale=0.35]{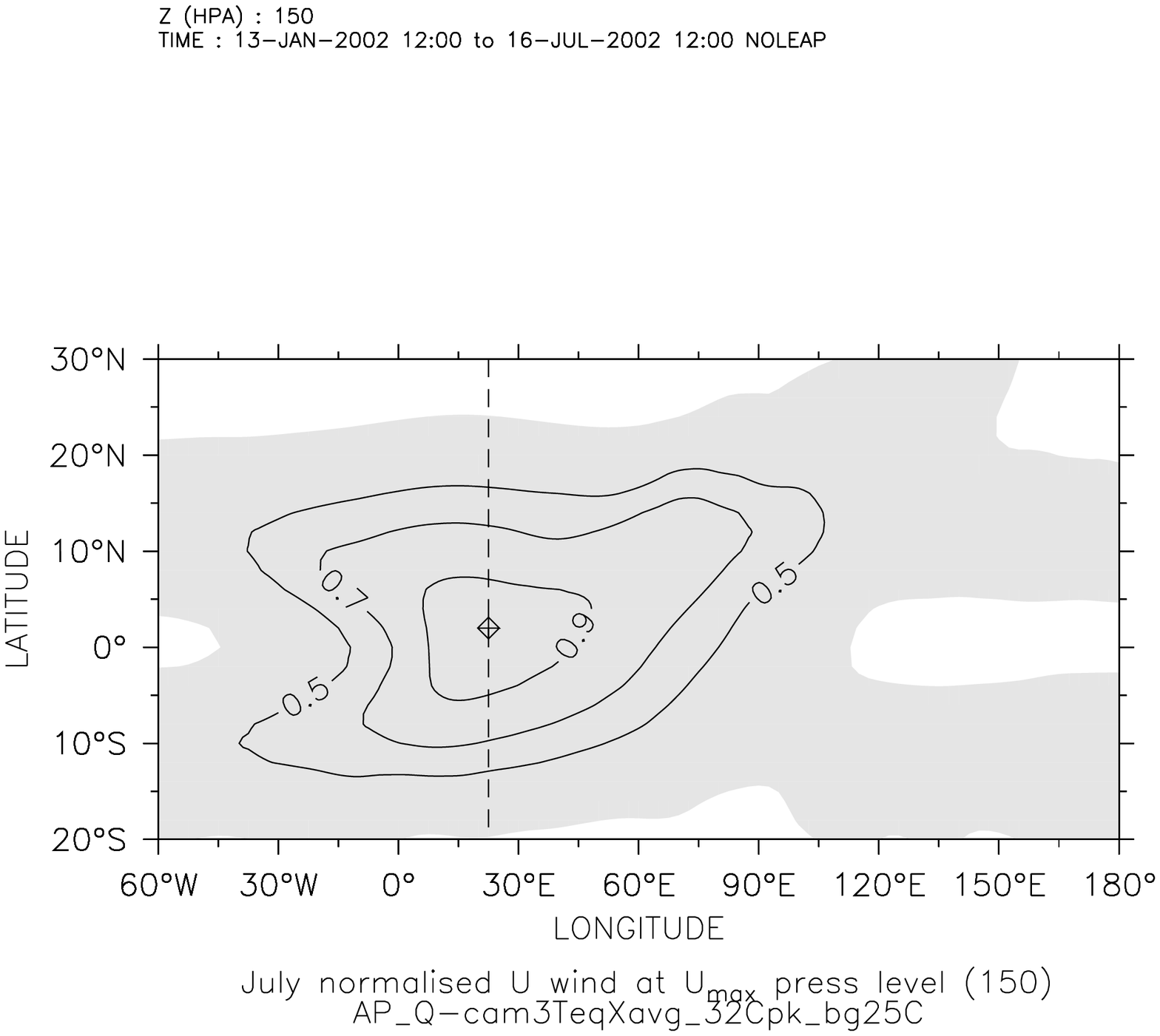}}
   \hspace{5mm}
   \subfloat[\nglo{}\_ip\_nf: \ip{}]{\label{nglo-cam3TeqXavg-Unorm-xy}\includegraphics[trim = 5mm 20mm 10mm 60mm, clip, scale=0.35]{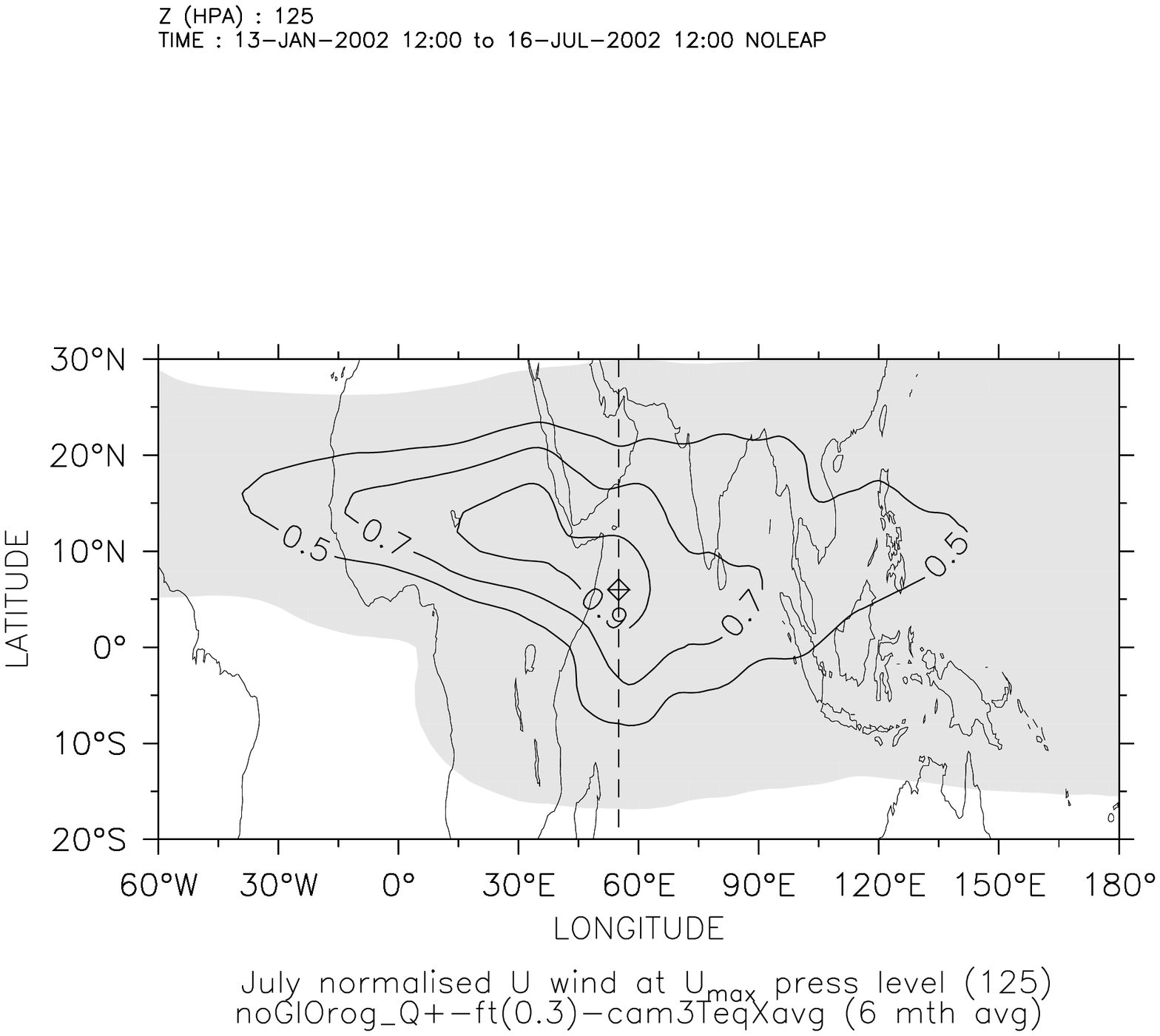}}
   \vskip 5mm
   \subfloat[AP\_90e20n\_ip\_pf: \ip{}]{\label{ap-ft0.3-cam3TeqXavg-Unorm-3225-xy}\includegraphics[trim = 5mm 20mm 10mm 60mm, clip, scale=0.35]{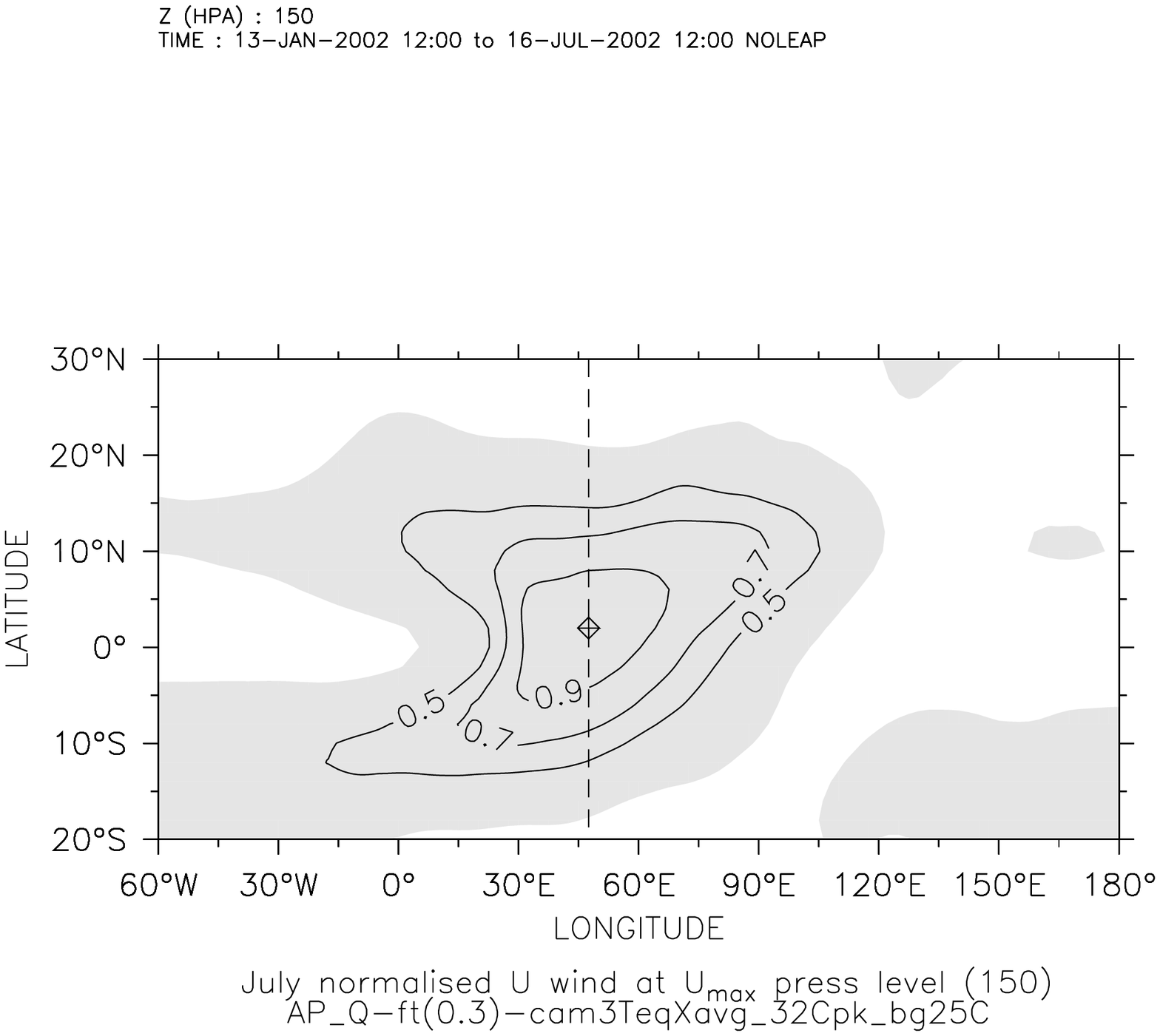}}
   \hspace{5mm}
   \subfloat[\nglo{}\_ip\_pf: \ip{}]{\label{nglo-ft0.3-cam3TeqXavg-Unorm-xy}\includegraphics[trim = 5mm 20mm 10mm 60mm, clip, scale=0.35]{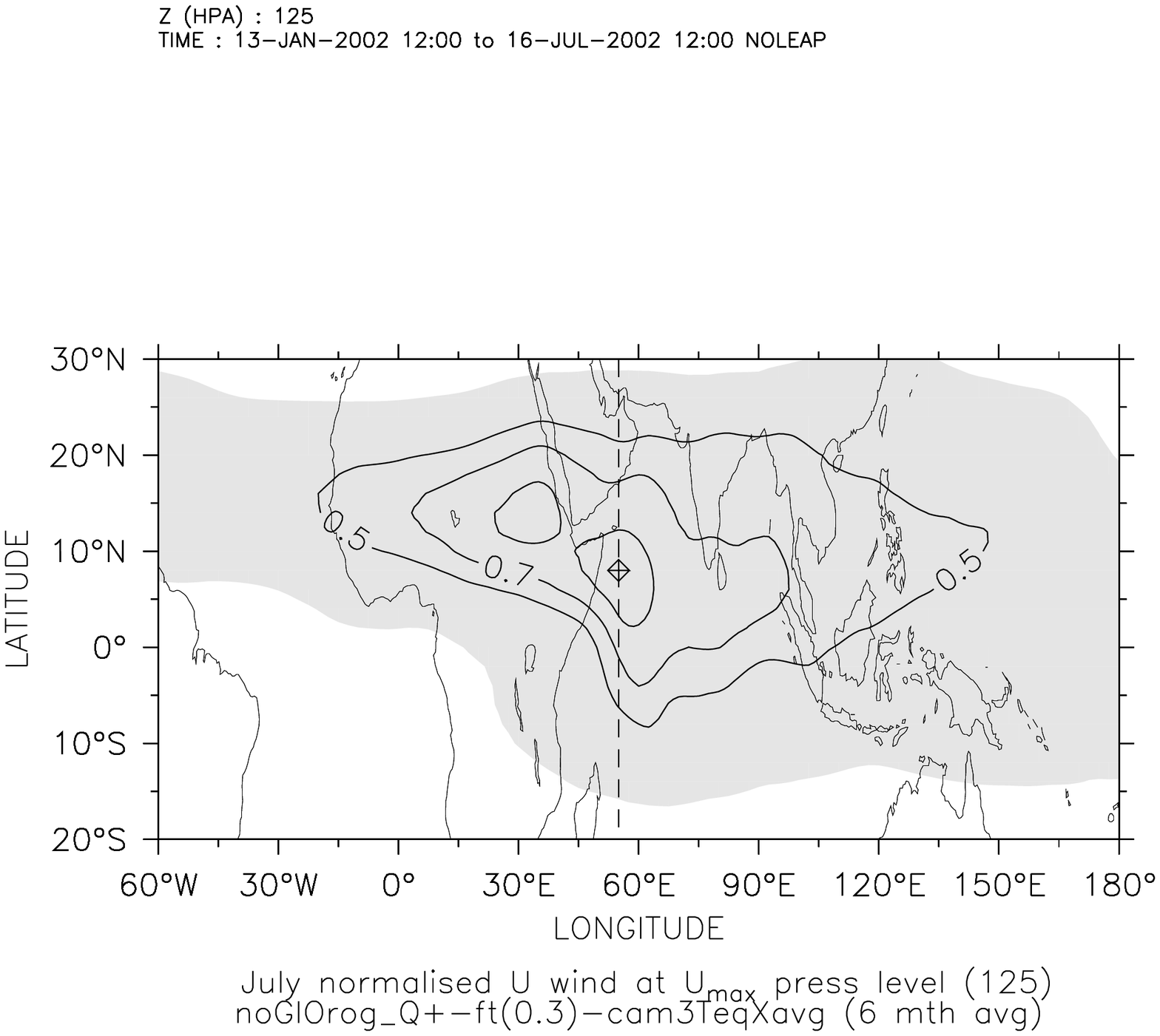}}
   \end{center}
   \caption{\un{} contours. Left panel is AP\_90e20n simulation set, right panel is \nglo{} simulation set. The \ip{} simulations have heating and zonally averaged equilibrium temperatures from the corresponding \fp{} simulation. Cross-sections are at \um{} pressure level, `cross-diamond' is location of \um{}. Grey region denotes easterlies. Additional details are in Table \ref{tab: sims} and Fig. \ref{fig: nglo-ap-qmax-xy}.}
   \label{fig: ap-nglo-q-Unorm-xy}
\end{figure}

For comparison purposes the \un{} contours of actual \ap{} simulation of the 12$^{th}$ month is shown in Fig. \ref{ap-Unorm-3225-xy}. For the \ip{} simulations, the differences were found to be minor when either temperatures at each grid point or zonally averaged temperatures were specified. Hence we only show the latter. This lack of difference demonstrates that provided radiative-convective equilibrium temperatures are prescribed, heating plays a dominant role in determining the profile of the TEJ.

\begin{table}[htbp]
\caption{Magnitude (\ms{}) and location of peak zonal wind (\um{})}
\label{tab: sims}
\begin{center}
\begin{tabular}{ccccccc} \toprule
\mr{2}{*}{Case}    & \mr{2}{*}{Simulation type} & Upper tropospheric & \mc{4}{c}{Zonal wind}                 \\ \cline{4-7}
                   &                            & friction           & Peak  & Lon       & Lat         & Press (hPa) \\ \midrule
AP\_90e20n         & \fp{}                      & \cam{}             & 41.64 & 47.5\de{} & 4\dn{}      & 150 \\
AP\_90e20n\_ip\_nf & \ip{}                      & Absent             & 42.35 & 27.5\de{} & 0\degrees{} & 150 \\
AP\_90e20n\_ip\_pf & \ip{}                      & Present            & 39.36 & 47.5\de{} & 2\dn{}      & 150 \\ \midrule
\nglo{}            & \fp{}                      & \cam{}             & 42.34 & 30.0\de{} & 10\dn{}     & 125 \\
\nglo{}\_ip\_nf    & \ip{}                      & Absent             & 46.09 & 55.0\de{} & 6\dn{}      & 125 \\
\nglo{}\_ip\_pf    & \ip{}                      & Present            & 44.09 & 55.0\de{} & 8\dn{}      & 125 \\ \bottomrule
\mc{7}{c}{AP\_90e20n: \ap{}; \nglo{}: no orography} \\ \bottomrule
\end{tabular}
\end{center}
\end{table}

In the absence of friction, it is seen from Fig. \ref{ap-cam3TeqXavg-Unorm-3225-xy} that the zonal location of \um{} is eastwards by 20\degrees{} compared to \fp{} simulation (\ref{ap-Unorm-3225-xy}). However, use of \ut{} causes the location of \um{} (Fig. \ref{ap-ft0.3-cam3TeqXavg-Unorm-3225-xy}) to be very similar to \fp{} simulation. Table \ref{tab: sims} confirms this. Similar trends were also observed in the meridional profiles of \un{}. The zonal profile with friction was also more realistic compared to the simulation without friction.

Though not shown, \ip{} simulations were also conducted using zonally averaged temperatures from \nglo{} simulation described in section \ref{intro}. In the absence of friction the jet in \ip{} became more zonally extensive (Fig. \ref{ap-cam3TeqXavg-Unorm-3225-xy}). But presence of \ut{} made the zonal profile similar to Fig. \ref{ap-ft0.3-cam3TeqXavg-Unorm-3225-xy} . This can be understood from the thermal wind equation (write it). The zonal thermal wind ($u_T$) is proportional to meridional temperature gradient as follows: $u_T = \delta{}u_g \approx \frac{R_a}{fP}\left(\frac{\partial{}T}{\partial{}y}\right)_P(\delta{}P)$, where, $R_a$ is gas constant for dry air, $f$ is Coriolis parameter, $P$ is pressure and $T$ is temperature. Compared to \ap{}, heating in \nglo{} is elevated over a significant zonal stretch (refer Fig. \ref{nglo-qmax-xy}). Thus using zonally averaged temperatures from \nglo{} implies that temperatures are on an average elevated in the zonal direction in comparison to \ap{} simulation. Thus meridional temperature gradients do not change over a significant zonal scale. This implies a tendency for the zonal wind to be uniform over a greater zonal span. In the absence of \ut{} this leads to a significant increase in jet length. With the location of \um{} similar to Fig. \ref{ap-cam3TeqXavg-Unorm-3225-xy}. But friction again caused the magnitude of \um{} and zonal wind pattern to be similar to Fig. \ref{ap-ft0.3-cam3TeqXavg-Unorm-3225-xy}. This further shows the importance of \ut{}.

\subsection{\nglo{} simulation} \label{nororg}

The heating imposed in the \ip{} simulations is from July 1$^{st}$ year \fp{} simulation. The heating is shown in Fig. \ref{nglo-qmax-xy}. Fig. \ref{nglo-Unorm-xy} shows the horizontal cross-section of \fp{} simulation. As with the \ip{} \ap{} simulations, \ip{} simulations were conducted with zonally averaged temperature profiles and temperature specified at each grid point. Since was no major difference in the structure of the TEJ between the two sets, only the former cases are shown.

From Figs. \ref{nglo-cam3TeqXavg-Unorm-xy},\subref*{nglo-ft0.3-cam3TeqXavg-Unorm-xy} we can see that \ip{} simulation has correctly reproduced the horizontal structure of the TEJ.  Although the location of \um{} is  much westwards compared to \fp{}, there is a broad region of zonal wind maxima. The velocity differences in the region enclosed by the 0.9 contour was found to be about 2-3\ms{}. This region is more zonally extensive than the \fp{} simulation. It is also seen that \ut{} does not have much influence on the simulation except that it slightly reduces the value of \um{} (refer Table \ref{tab: sims}). This further implies that when, without \ut{}, the \ip{} simulation is realistic, presence of friction does not make it unrealistic.

\section{The Tropical Easterly Jet in Gill model and its comparison with \cam{}}

\subsection{Zonal wind from Gill model} \label{g80}

There has been a lack of interest in studying the characteristics of the Tropical Easterly Jet. This is partly due to the remarkable paper by \cite{gill-80} which is a linear shallow water model using the equatorial $\beta$-plane approximation. Although critical analysis and criticisms (for example \cite{neelin-89}, \cite{plumb-10}) for this model have been given, it still remains a pioneering work.

The asymmetric solution of Gill's formulation has been considered. The analytical expressions for heating and the resultant zonal winds are given by Eqs. \eqref{gill-q} and \eqref{gill-u}.

\begin{equation}\label{gill-q}
Q(x,y) = F(x)y\exp{}\left(-\frac{y^2}{4}\right)
\end{equation}
where, \\
\[
F(x) =
\begin{cases}
   \cos{}(kx) & \text{if } \lvert{}x\rvert{} < L \\
   0 & \text{if } \lvert{}x\rvert{} > L
\end{cases}
\]
\[
   k = \frac{\pi{}}{2L}
\]

\begin{equation}\label{gill-u}
u(x,y) = \frac{1}{2}q_3(\epsilon{},x)(y^3-6y)\exp{}\left(-\frac{y^2}{4}\right)
\end{equation}
where, \\
\[
q_3(\epsilon{},x) =
\begin{cases}
   -k\big[1+\exp{}\left(-10\epsilon{}L\right)\big]\exp{}\left\{5\epsilon{}(x+L)\right\} & \text{if } x < -L \\
   -5\epsilon{}\cos{}(kx)+k\big[\sin{}(kx)-\exp{}\left\{5\epsilon{}(x-L)\right\}\big] & \text{if } \lvert{}x\rvert{} < L \\
   0 & \text{if } x > L
\end{cases}
\]
$\epsilon{}$ = 0.1, a factor used to represent Rayleigh friction and Newtonian cooling, \\
$L$: non-dimensional zonal length of the domain across which forcing is considered

\begin{align*}
\begin{cases}
Q_{max} = +\sqrt{2}\text{e}^{-\frac{1}{2}} \text{ at (x,y) = } \left(0,\sqrt{2}\right) \\
Q_{min} = -\sqrt{2}\text{e}^{-\frac{1}{2}} \text{ at (x,y) = } \left(0,-\sqrt{2}\right)
\end{cases}
\end{align*}

The heating and zonal wind reach their extremum values at the locations given below:

\begin{align*}
\begin{cases}
u_{min} = -2.45 \text{ at (x,y) = } \left(-1.14,\left(6-2\sqrt{6}\right)^\frac{1}{2}\right) \\
u_{max} = +2.45 \text{ at (x,y) = } \left(-1.14,-\left(6-2\sqrt{6}\right)^\frac{1}{2}\right)
\end{cases}
\end{align*}

The three-dimensional structure of heating is obtained by multiplying Eqs. \eqref{gill-q} and \eqref{gill-u} as follows:

\begin{subequations}\label{gill-dim}
\begin{align}
Q(x,y,z) = Q(x,y)\sin{}\left(\frac{\pi{}z}{D}\right)\label{gill-dim-q} \\
u(x,y,z) = u(x,y)\cos{}\left(\frac{\pi{}z}{D}\right)\label{gill-dim-u}
\end{align}
\end{subequations}
where, $0 \leq{} z \leq{} D$, $D$ being the non-dimensional height of the domain

Using the standard formulae given in G80, the equations have been non-dimensionalized as follows:

\begin{gather*}
c = \sqrt{gH} , L_{d} = \sqrt{\frac{c}{2\beta{}}}, \tau{}_{d} = \frac{L_{d}}{c} = \sqrt{\frac{1}{2\beta{}c}}, \epsilon{}_{d} = \frac{\epsilon{}}{\tau{}_{d}}
\end{gather*}
where, $g$ = 9.81 m s$^{-2}$, $H$ = 400 m, $c$ = 62.64\ms{} (speed of gravity waves in the absence of rotation), $L_{d}$ = 10\degrees{} (horizontal length scale), \td{} = 4.93 hour (time scale), $\epsilon{}_{d}$ = 0.021 hour$^{-1}$ (the dimensionalized dissipation rate). This then gives the heating to be dimensionalized by $Q_{d}$=\td$^{-1}$ = 4.87\kd{}.

Additional details are given in Table \ref{tab: gill_90e15n_Umax}. The heating maxima is centered at 0\degrees{},10\dn{}. The meridional and horizontal structure of the normalized zonal winds is shown in Figs. \ref{gill-0.1-Unorm-yz} and \ref{gill-0.1-Qmax,Unorm-xy} respectively. The default value of $\epsilon$, 0.1, gives very high zonal winds speeds. For this case the zonal distance of \um{} is $\sim$11.5\degrees{} to the west of the heating maxima. Zonal wind speeds become reslistic when the value of $\epsilon$ is increased, for example, to 1. For $\epsilon$ = 1, the zonal distance of \um{} is just 2\degrees{} to the west of heating maxima. The jet is also more zonally confined and appears symmetric about the peak. Thus, reducing zonal wind speed has put a penalty on the location and shape of the jet. The meridional shape is independent of $\epsilon$ and is thus same as Fig. \ref{gill-0.1-Unorm-yz}. Zonal half length and width listed in Table \ref{tab: gill_90e15n_Umax} show a very short jet for standard and high $\epsilon$ values.

\begin{table}[htbp]
\caption{Maximum heating rates, magnitude and location peak easterly, and jet half-dimension}
\label{tab: gill_90e15n_Umax}
\begin{center}
\begin{tabular}{cccccccc} \toprule
\mc{7}{c}{Gill model *} \\ \midrule
\mr{2}{*}{$\epsilon{}$}            & $Q_{max}$       & \mc{4}{c}{\textit{\um{} (\ms{})}}               & \mc{2}{c}{\textit{Jet dimension}} \\ \cline{3-8}
                                   & (\kd{})         & Value  & Lon       & Lat                & Level & Length          & Width \\ \midrule
0.1                                & \mr{2}{*}{5.67} & 153.48 & 11.4\dw{} & \mr{2}{*}{10.5\dn} & Top   & 34.7\degrees{}  & \mr{2}{*}{15.3\dn}\\
1                                  &                 & 24.15  & 2.0\dw{}  &                    & Top   & 26.7\degrees{}  & \\ \midrule \midrule
\mc{7}{c}{Gill heating in \cam{} **} \\ \midrule
\mr{2}{*}{\textit{Case}}           & $Q_{max}$       & \mc{4}{c}{\textit{\um{} (\ms{})}}             & \mr{2}{*}{\textit{Jet width}}\\ \cline{3-6}
                                   & (\kd{})         & Value & Lon        & Lat                & Press (hPa) & \\ \midrule
Qgill\_ip\_noTeq                   & \mr{3}{*}{5.67} & 29.87 & 65\de{}    & 6\dn{}             & 175 & 25.1\degrees{} \\
Qgill\_ip\_APTeq                   &                 & 25.66 & 52.5\de{}  & 6\dn{}             & 225 & 28.2\degrees{} \\
Qgill\_ip\_nGlOTeq                 &                 & 31.16 & 72.5\de{}  & 6\dn{}             & 175 & 19.4\degrees{} \\ \bottomrule
\mc{8}{c}{* For Gill model length and heating have been dimensionalized by 10\degrees{} and 4.87\kd{}} \\
\mc{8}{c}{**  For Gill heating in \cam{}, the 0.5\um{} contour does not close, so jet does not exist}
\end{tabular}
\end{center}
\end{table}

\subsection{Gill heating in \ip{} version of \cam{}} \label{gcam}

It is generally assumed that the TEJ can be explained with the help of Gill's formulation. But this is quite a simplistic viewpoint since the tropospheric structure has significant diffrences in upper and lower halves. More importantly from a dynamical viewpoint, friction is not fully understood and hence its parametrization differs significantly in both layers. As was shown before changing friction in G80 itself changes the zonal wind pattern.

The advantage of using an AGCM is that all the necessary non-linearities in the governing dynamical equations is present. This along with multiple layers in the vertical will reveal the response to tropical forcing which is closer to reality.

\begin{figure}[htbp]
   \begin{center}
   \subfloat[Gill's solution: \un{}]{\label{gill-0.1-Unorm-yz}\includegraphics[trim = 0mm 15mm 80mm 65mm, clip, scale=0.35]{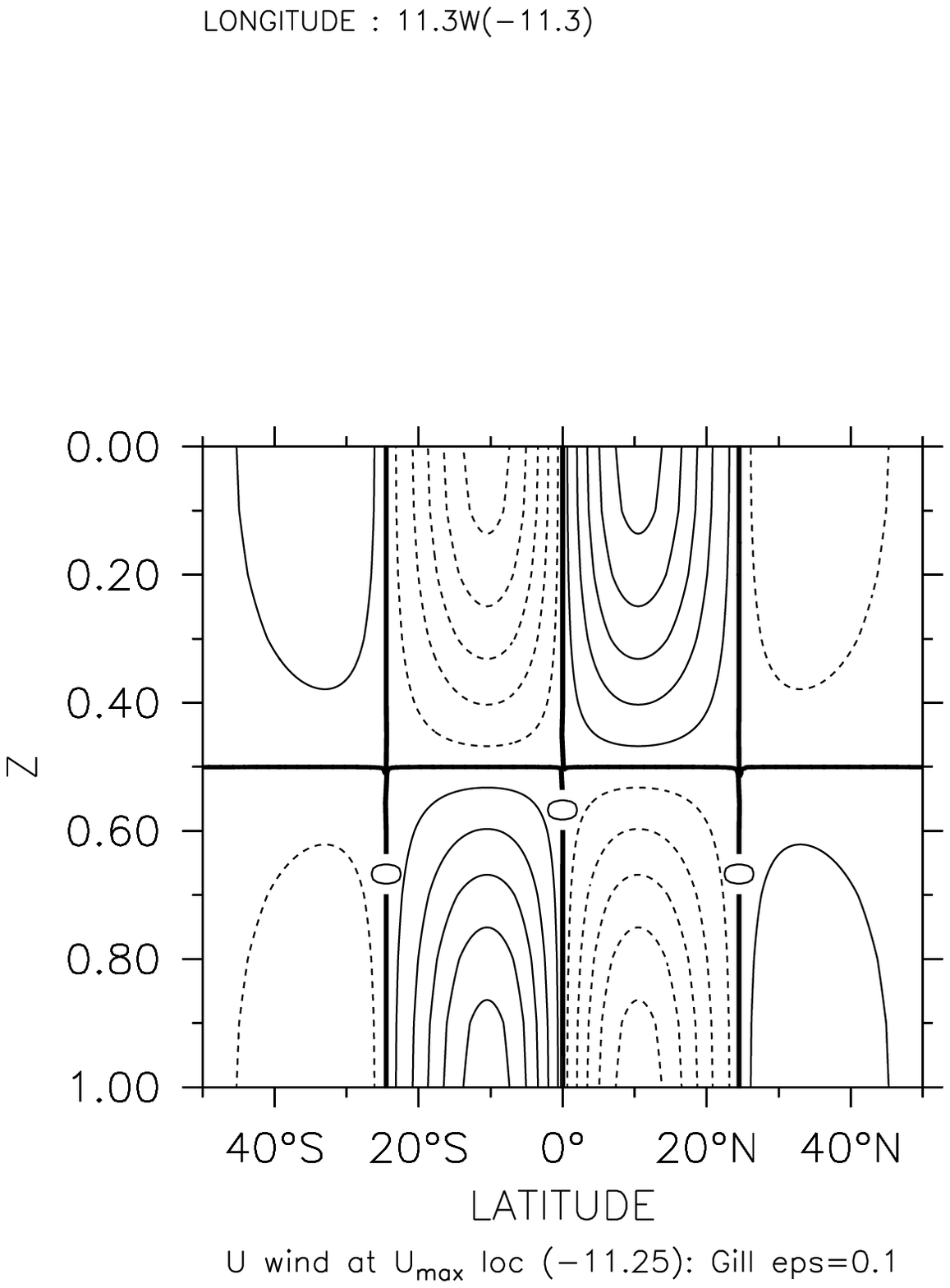}}
   \hspace{5mm}
   \subfloat[Gill's solution: heating rate (red contours) and \un{}; $\epsilon$=0.1, \um{}=153.48\ms{}]{\label{gill-0.1-Qmax,Unorm-xy}\includegraphics[trim = 5mm 20mm 10mm 60mm, clip, scale=0.35]{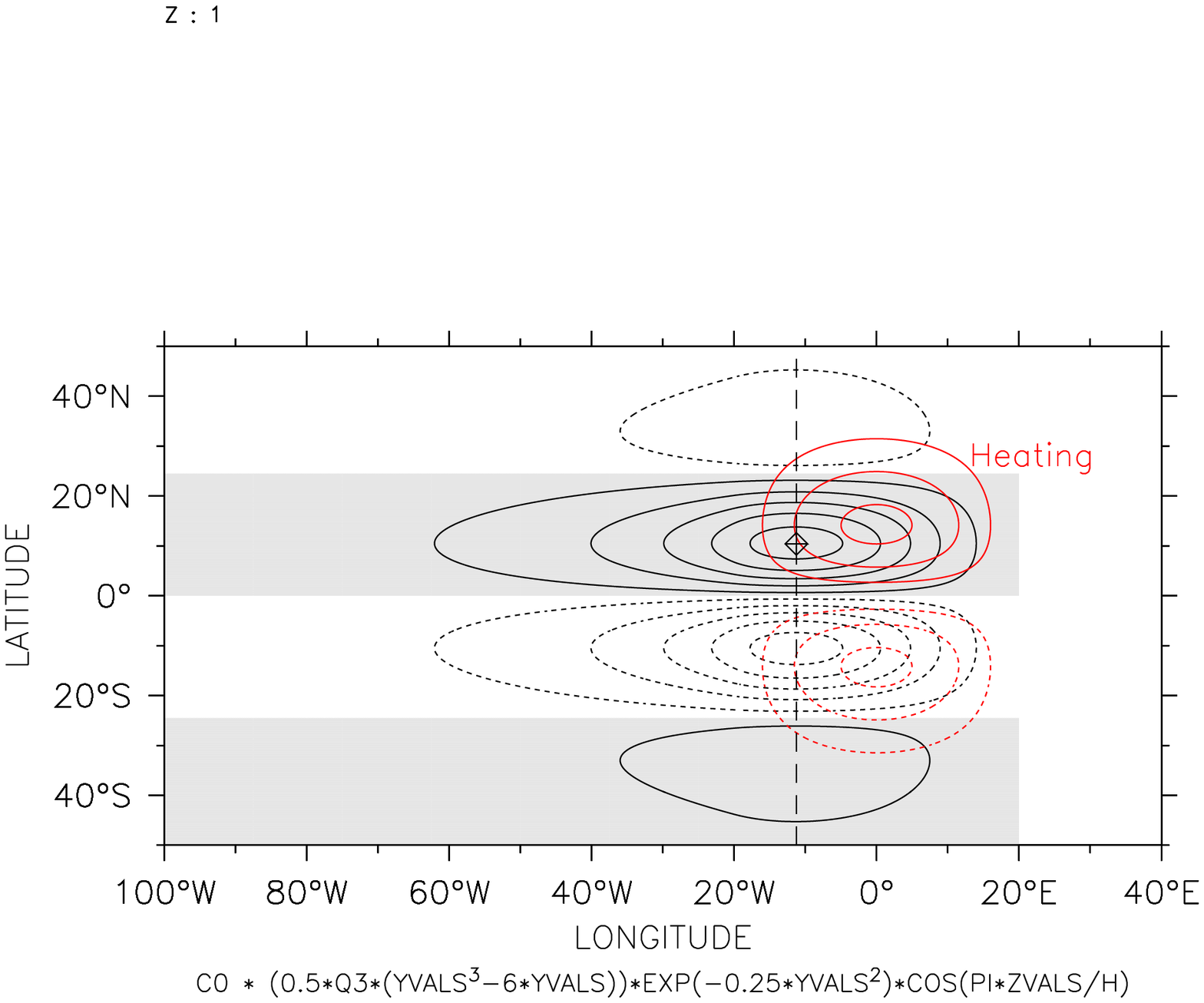}}
   \vskip 5mm
   \subfloat[\cam{}, \ip{}: \un{}]{\label{gill-idl-z2degTeqXavg-Unorm-yz}\includegraphics[trim = 0mm 15mm 80mm 65mm, clip, scale=0.35]{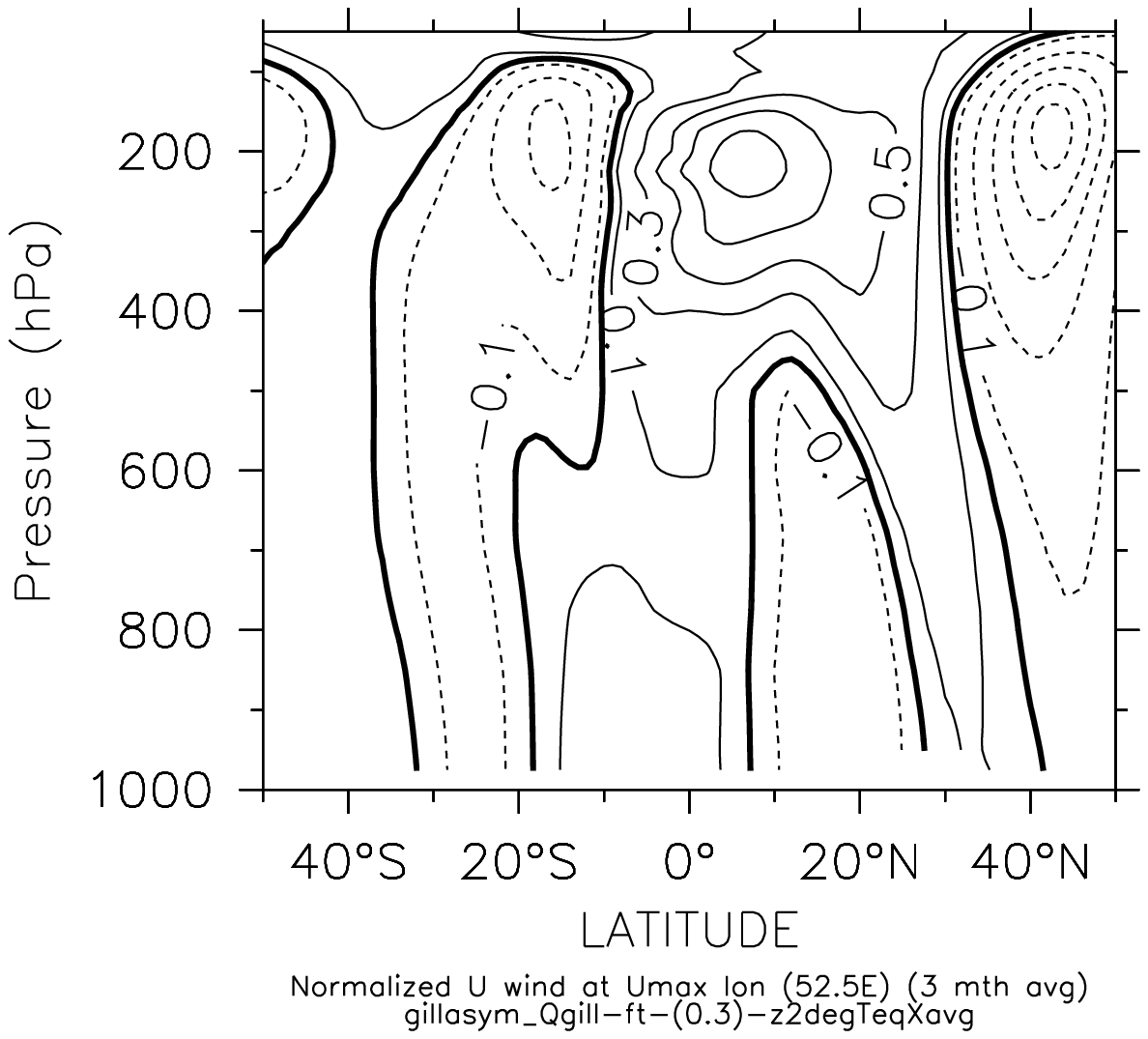}}
   \hspace{5mm}
   \subfloat[\textit{Ideal-physics}: heating rate (red contours) and \un{}]{\label{gill-idl-z2degTeqXavg-Unorm-xy}\includegraphics[trim = 5mm 20mm 1mm 60mm, clip, scale=0.35]{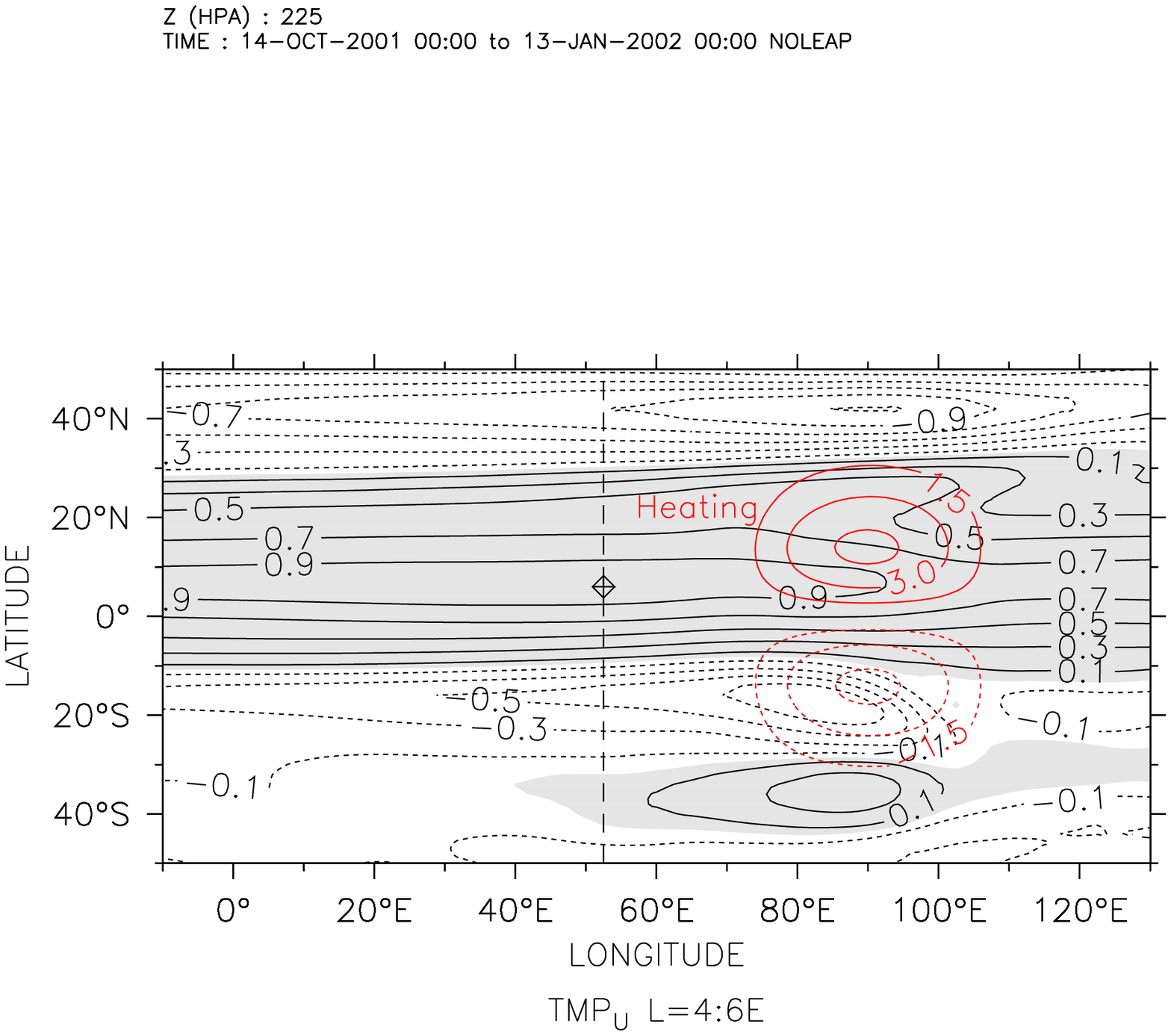}}
   \end{center}
   \caption{\protect\subref{gill-0.1-Unorm-yz},\protect\subref{gill-0.1-Qmax,Unorm-xy} Gill's solution: \un{} contours at \um{} zonal location; \protect\subref{gill-idl-z2degTeqXavg-Unorm-yz},\protect\subref{gill-idl-z2degTeqXavg-Unorm-xy} \ip{} simulation with analytical temperature profile (APTeq, refer Table \ref{tab: gill_90e15n_Umax}) imposed. Heating rates are in red (contour interval 1.5\kd{}, negative values are dashed), black lines denote \un{} contours, `cross-diamond' shows the location of maximum easterly wind. Grey region denotes easterlies.}
   \label{fig: gill-ip,U-xy-yz}
\end{figure}

The same heating profile discussed in section \ref{g80} is imposed in the \ip{} version of \cam{}. The peak of zonal heating has has been shifted to 90\de{} instead of being at the origin as in the Gill model. This has been done to facilitate easy comparison with the \ap{} simulations discussed in section \ref{aqua}. Due to zonal symmetry there is be no dynamic change had the source been kept at Greenwich meridian. Heating maximizes at 500hPa and becomes zero at the lowest model level and tropopause (in \cam{} the tropopause pressure level, $TP_p$, has been defined as $TP_p=250-150\cos{}\phi$, where $\phi$ is the latitude). Lower tropospheric friction has been retained as before, that is, it is the same as HS94. Since \ut{} has been found to be necessary, the previously described Rayleigh friction parametrization has been used.

Two different temperature profiles have been imposed. The first profile was obtained by zonally and meridionally averaging the 12$^{th}$ month of AP\_90e20n simulation discussed before. Thus a single set of points was obtained at each model level. The surface temperature was found to be close to 298K. The tropopause temperature was fixed as 205K. A second degree polynomial was used to fit the values between the lowest model level to the tropopause. Beyond tropopause a linear fit was used till topmost model level which was close to 235K. This profile thus has no meridional asymmetry and the simulation is named Qgill\_ip\_APTeq. The second profile used had zonally averaged temperatures obtained from July 1$^{st}$ year \nglo{} simulation. This profile serves to understand the zonal wind response in the presence of meridionally asymmetric temperature profiles which actually exist during the period when TEJ is observed. This latter simulation is named Qgill\_ip\_nGlOTeq. Additionally a set of simulations without any temperatures imposed has also been carried out. This is named as Qgill\_ip\_noTeq case. This last set is unrealistic and has been done to know the response to heating only. The initial conditions are the same as the \ap{} \fp{} simulations.

The peak easterly zonal wind and its location for all simulations are listed in Table \ref{tab: gill_90e15n_Umax}. Normalized zonal wind profiles at the location of peak zonal wind are shown in Figs. \ref{gill-idl-z2degTeqXavg-Unorm-yz},\subref*{gill-idl-z2degTeqXavg-Unorm-xy}. The meridional section shows that \cam{} qualitatively captures the Gill model features. But the horizontal section shows extensive jet lengths. In fact as documented in Table \ref{tab: gill_90e15n_Umax}, the jet is undefined in the \ip{} simulations. The easterlies extend to the south of the Equator and have greater width compared to westerlies. Thus the \ip{} \cam{} simulation is far from being meridionally symmetric despite heating being symmetric. The easterlies also have their maxima between 5\dn{}-10\dn{}. Westerlies peak aroundf 20\ds{}. The meridional cross-section reveals that the asymmetry exists at all vertical levels. The low-level winds are sometimes four times weaker than those at upper levels.

In terms of jet magnitude, the value of \um{} is in close agreement with Gill solution for $\epsilon$ = 1 which represents very high damping in the upper troposphere. The location of \um{} does not match. The zonal distance between \um{} and heating is between 20\degrees{}-30\degrees{}.

These simulations thus clearly show that the Gill model though very useful is unable to capture the true vertical zonal wind structure revealed in a multi-layered model with non-linear interactions and more complicated frictional parametrization. It can be expected the when multiple heat sources are considered these inadequacies will be further exposed. Further, when full AGCM physics is taken into account, there will be further deviations from the Gill model. Beyond crucial and insightful qualitative explanations, the analytical formulation of Gill cannot give accurate quantitative answers.

\section{Conclusions}

An AGCM, \cam{}, has been run with idealized physics and the simulation has been compared with the corresponding \fp{} simulations. The \ip{} simulations used a very simple frictional parametrization. In the absence of \ut{}, using the same heating from \fp{} simulation caused the jet to be incorrectly simulated. Incorporating \ut{} made the jet closely match the corresponding \fp{} simulation.

The Gill model predicts the presence of a TEJ. But unless very high friction is used, the magnitude of zonal winds is very high. Heating from Gill model has been applied in \cam{} to check if an AGCM can reproduce a TEJ. The \ut{} friction which gives realistic simulation in \cam{} has been used. It has been demonstrated that Gill heating does not lead to the formation of a TEJ.

\bibliographystyle{usrdef}
\bibliography{references}

\end{document}